\documentstyle[12pt]{article}

\def\abstract#1{\begin{center}{\large ABSTRACT}\end{center} \par #1}
\def\title#1{\begin{center}{\Large\bf {#1}}\end{center}}
\def\author#1{\begin{center}{\large #1}\end{center}}
\def\address#1{\begin{center}{\it #1}\end{center}}

\begin{document}
\begin{titlepage}
\hspace*{\fill}
\vbox{
 \hbox{UT-Komaba/97-6}
  \hbox{hep-th/9703070}
  \hbox{March 1997}}
\vspace*{\fill}
\begin{center}
  \Large\bf
Stringy Cosmic Strings \\
and Compactifications of F-theory 
\end{center}

\vskip 1cm
\author{
  Masako Asano \footnote{E-mail address: \tt asano@hep1.c.u-tokyo.ac.jp}} 
\address{
Institute of Physics,
  University of Tokyo,\\
  Komaba, Meguro, Tokyo 153, Japan
}
\vspace{\fill}
\abstract{
  We construct stringy cosmic string solutions corresponding to 
compactifications of F-theory on several elliptic Calabi-Yau manifolds
by solving the equations of motion of low energy effective 
action of ten dimensional type IIB superstring theory.
Existence of such solutions 
supports the 
compactifications of F-theory.
} \vspace*{\fill}
\end{titlepage}
\section{Introduction}
Recently, `string dualities' among various superstring theories
including M- and F-theory 
have been greatly investigated and 
several phenomena related to non-perturbative behavior of string theory 
have been found~\cite{vare}.
F-theory~\cite{va} is considered as a twelve dimensional theory underlying 
the $SL(2,{\bf Z})$ symmetry of ten dimensional type IIB theory.
This theory may be considered as a key 
ingredient of the `duality web,' 
though the corresponding microscopic theory has not been constructed.

Compactification of F-theory on 
an $n$-dimensional Calabi-Yau manifold
(Calabi-Yau $n$-fold) $X$ 
makes sense if $X$ admits an elliptic fibration over some
$(n\!-\!1)$-dimensional base $B$.
It is defined as compactification of type IIB theory on $B$
with 
the field $\tau\equiv \tilde{ \phi} + i e^{-\phi}$
varies on $B$, i.e., $\tau= \tau(z^i)$
($\{z^i\}\in B$).
Here $\phi$ is the dilaton, $\tilde{\phi}$ is the scalar in 
the R-R sector and $\tau$ 
is identified with the complex structure modulus of the fiber
torus.

For example, consider a compactification of F-theory on $K3$ surface which 
admits elliptic fibration over the base ${\bf CP}^1$.
In terms of type IIB theory, this model is interpreted as a compactification
of ten dimensional type IIB theory 
compactified on ${\bf CP}^1$ with 24 D 7-branes~\cite{pol} 
located on it.
We have so called stringy cosmic string solutions of type IIB theory 
such that ten dimensional spacetime is ${\bf CP}^1\times {\cal M}^8$
and generically on 24 points in ${\bf CP}^1$ the field
$\tau$ diverges.
On the other hand, from the F-theory point of view, 
we understand this property of $\tau$ from 
the geometrical structure of elliptic $K3$ surface.
Moreover, 
the structure of  moduli space of the theory
such as symmetry enhancement or massless spectrum can be deduced 
only from the geometric data.
The resulting eight dimensional theory is conjectured to be dual to 
heterotic string theory compactified on torus.
One of the evidences of this duality is 
that \cite{sen} there is a relation between F-theory
on $K3$ near the orbifold limit of $K3$
and T-dual of type I theory on torus 
(type I$'$ orientifold theory).

Compactifications of F-theory to lower dimensions have also been 
investigated so far\cite{mv,ber,bj}.
In particular, six dimensional theories obtained by compactifications
on several elliptic Calabi-Yau three-folds are investigated 
well.
It is known that a large class of $d\!=\!6$, $N\!=\!1$ superstring theories 
can be represented by F-theory compactification and 
they are conjectured to be dual to other string theories.
As for compactifications on Calabi-Yau four-folds, 
the existence of non-perturbative superpotential
of the four dimensional theory is determined by geometry \cite{wi}.
In this way, F-theory compactification on 
Calabi-Yau 3,4-folds seems to be succeeded in itself.
However, 
evidences such as 
stringy cosmic string solutions~\cite{gsvy}
corresponding to these compactifications are still lacking.
In order to justify the F-theory compactification, 
it seems that there must be such solutions
as in the case of eight dimensions, $F/K3$,
since the compactification
is originally defined as type IIB compactification with 
certain properties.

Furthermore, if there exist a stringy cosmic string 
solution representing
an F-theory compactification on a Calabi-Yau manifold, it
gives  explicit metric on the base manifold $B$ and 
in a certain limit it
is important in identifying the base manifold with, e.g., an orbifold.

In this article, we search for stringy cosmic string solutions 
corresponding to various F-theory compactifications by
extending the method in ref.\cite{gsvy}.
We obtain several solutions which are considered
to represent F-theory on elliptic Calabi-Yau manifold 
with $B={\bf CP}^1\!\times\! {\bf CP}^1$ and with the Hirzebruch 
surface ${\bf F}_n$,
and then we analyze their constant coupling limit.
We also consider the compactification on Calabi-Yau three-fold
with hodge numbers $h^{1,1}\!=\!h^{1,2}\!=\!19$ described in ref.\cite{va}.

This paper is organized as follows:
First, we explain general properties of F-theory compactifications
in section~2 
and stringy cosmic string solutions corresponding to 
F-theory compactification on $K3$ in section~3.
In section~4, extending the argument of section 3,
we obtain stringy cosmic string solutions 
corresponding to F-theory compactifications on 
several elliptic Calabi-Yau 3,4-folds and 
consider their constant coupling limit.
In section~5, we give conclusions briefly.

\section{Compactifications of F-theory}
F-theory is considered as a twelve dimensional theory underlying 
the conjectured  $SL(2,{\bf Z})$ duality
symmetry of type IIB theory in ten dimensions.
Ten dimensional type IIB theory contains $g_{\mu\nu}$, $B^A_{\mu\nu}$, $\phi$ 
(NS-NS fields) 
and $B^P_{\mu\nu}$, $\tilde{ \phi}$, $A^+_{\mu\nu\rho\sigma}$ (R-R fields) 
as bosonic fields in the low energy effective field theory.
The field $\tau=\tilde{ \phi} + i \exp(-\phi)$ transforms as
the complex modulus of a torus under an $SL(2,{\bf Z})$
duality transformation.

Compactification of F-theory to lower dimensions is formulated as
that of type IIB theory on a manifold $B$
with $\tau$ varies according to a position in $B$~\cite{va}.
More precisely, compactification of F-theory on a manifold $X$
which admits elliptic fibration over a base $B$ is 
type IIB compactification on 
$B$ with $\tau=\tau(z^i)$ where $\{z^i\}\in B$.

In the case of elliptic $K3$, base manifold is ${\bf CP}^1$
and the elliptic fiber is given on each $z\in {\bf CP}^1$ in the
Weierstrass form :
\begin{equation}
y^2=x^3+f_8(z)x+g_{12}(z)
\end{equation}
where $z$ is affine coordinate of ${\bf CP}^1$, and 
$f_8$ and $g_{12}$ are polynomials of degree 
$\le 8$
and $\le 12$ in $z$ respectively.
The elliptic fiber degenerates 
when the discriminant
\begin{equation}
\Delta=4 f^3+ 27 g^2
\end{equation}
vanishes, which happens generically 24 points on ${\bf CP}^1$.
In the point of view of type IIB theory, this phenomenon 
represents that there are 24
D 7-branes located on ${\bf CP}^1$ 
and each of them carries a magnetic charge for $\tilde{\phi}$. 
The modulus $\tau$ of the fiber torus is given as
\begin{equation}
J(\tau(z))=\frac{4 (24f)^3}{4f^3+27g^2}\, .
\label{jtau}
\end{equation}
Here $J(\tau)$ is the modular invariant function of $\tau$ defined as 
\begin{equation}
J(\tau)=\frac{(\theta_2(\tau)^8+\theta_3(\tau)^8+\theta_4(\tau)^8)^3}
{\eta(\tau)^{24}} \, .
\end{equation}
Geometry of elliptic $K3$ can teach us the spectrum 
of the resulting eight dimensional theory, and the theory
is conjectured to be dual to heterotic string theory compactified on 
$T^2$~\cite{va,sen}.

Similarly, we can define
compactifications of F-theory on a Calabi-Yau three-fold $X$
which admit elliptic fibration over a complex two dimensional 
base $B$ \cite{mv}.
If we choose $B$ suitably,
the resulting six dimensional theories have $N\!=\!1$ supersymmetry.
The number of tensor multiplets $T$, vector multiplets $V$
and hyper multiplets $H$ are related to the geometry of $X$ and 
$B$ as
\begin{eqnarray}
T &=& h^{1,1}(B)-1\\
r(V) &=& h^{1,1}(X)- h^{1,1}(B)-1\\
H &=& h^{2,1}+1
\end{eqnarray}
where $r(V)$ is the rank of the vector multiplet and
$V$ is determined by the singularity type of the fiber~\cite{mv,ber}.

For example, we can choose $X$ as an elliptic fibration over
the Hirzebruch surface ${\bf F}_n$, which is a ${\bf CP}^1$ bundle 
over ${\bf CP}^1$ characterized by an integer $n$.
The surface ${\bf F}_n$ is described as the quotient
\begin{equation}
(x,y,u,v)\sim (\lambda x, \lambda y, \mu u,\lambda^n\mu v)
\end{equation}
with $\lambda,\mu\in {\bf C}^\ast$
where $(x,y)$ and $(u,v)$ are considered as homogeneous coordinates 
of base ${\bf CP}^1$ and fiber ${\bf CP}^1$ respectively.
The elliptic fiber on ${\bf F}_n$ is represented as
\begin{equation}
y^2=x^3+\sum^{4}_{k=-4}f_{8-nk}(z')z^{4-k} x
+\sum^{6}_{k=-6}g_{12-nk}(z')z^{6-k}
\label{fn}
\end{equation}
where $z$ is affine coordinate on the fiber ${\bf CP}^1$
and $z'$ is that of the base  ${\bf CP}^1$.
Polynomials $f_{8-nk}$ and $g_{12-nk}$ are of degrees 
$\le 8-nk$ and $\le 12-nk$ respectively
and they are identical to zero when the coefficients are negative.
It can be seen from eq.(\ref{fn}) that the elliptic fiber degenerates 
on a codimension 1 surface $\Sigma$ in $B$. 
There are a number ($\le 24$) of connected components of $\Sigma$
and each of them is identified with
a part of D7-brane worldvolume.
These theories 
have one tensor multiplet
since $h^{1,1}({\bf F}_n)=2$.
In the case of $2\le n\le 12$,
it is conjectured that 
the theory is conjectured to be dual to 
compactification of $E_8\!\times\! E_8$
heterotic theory on $K3$ with instanton numbers 
$(12+n,12-n)$ for each $E_8$'s 
after higgsing as much as possible \cite{mv}.
In the case of $n=0$, i.e., $B={\bf CP}^1\!\times\!{\bf CP}^1$,
eq.(\ref{fn}) becomes
\begin{equation}
y^2=x^3+f(z,z') x+g(z,z')
\end{equation}
where f (or g) is of degree $\le 8$ (or $\le 12$) in each $z$ and $z'$
and this manifold represents the same Calabi-Yau manifold as in the
$n\!=\!2$ case. 

Other than ${\bf F}_n$, we have a large choice of $B$ 
in order to obtain six dimensional theories with $N\!=\!1$
supersymmetry~\cite{mv}. 

Among others we describe one example for future convenience.
That is an elliptically fibered Calabi-Yau manifold 
over ${\bf CP}^2$ blown up at nine points 
having hodge numbers
$h^{1,1}\!=\!h^{1,2}\!=\!19$.
This manifold is also represented as~\cite{va,mv} 
two elliptic fibrations over
${\bf CP}^1$ as 
\begin{equation}
y_i^2=x_i^3+f_i(z)x_i+g_i(z) \qquad (i=1,2)
\label{cy19}
\end{equation}
where $i\!=\!1,2$ represent two fiber tori on ${\bf CP}^1$
and $f_i$ (or $g_i$) is a polynomial of degree $\le 4$ (or $\le 6$).

\section{Type IIB stringy cosmic string solutions in eight dimensions}
One of the justifications of F-theory compactifications
on $K3$ is that there exist stringy cosmic strings~\cite{gsvy,ke}
of type IIB low energy effective theory that reflect 
main properties of elliptic $K3$.

The low energy effective action of type IIB theory in ten dimensions 
with  $B^A\!=\!B^P\!=\!A^+\equiv 0$
is 
\begin{equation}
S_{10}= -\frac 1 2\int d^{10}x \sqrt{g_{(10)}}
\left( R_{(10)} -\frac 1 2 \frac {\partial_\mu\tau g^{\mu\nu}\partial_\nu
{\bar \tau}}
{\tau_2^2} \right)
\label{action}
\end{equation}
where $\tau=\tilde{\phi}+i \exp(-\phi)(\equiv \tau_1+i\tau_2)$.
This action is invariant under an $SL(2,{\bf Z})$ modular transformation:
\begin{equation}
\tau\rightarrow \frac {a \tau +b}{c\tau +d} \quad , 
\quad
\left(
\begin{array}{ll}
a & b\\
c& d
\end{array}
\right) \in SL(2,{\bf Z}) .
\end{equation}
In order to obtain solutions of this action
corresponding to compactifications of 
F-theory on elliptic $K3$ over the base ${\bf CP}^1$, 
we set $\tau=\tau(z)$ where $z=x_8+ i x_9$ is affine coordinate on 
${\bf CP}^1$.
Correspondingly, we take an ansatz for ten dimensional metric as 
\begin{equation}
ds^2=e^{\phi(z,\bar{z})}dz d\bar{z} + dx^2_7 +\cdots +dx^2_1 -dx^2_0 .
\label{metric10}
\end{equation}
Then the equations of motion of the action (\ref{action}) are~\cite{gsvy}
\begin{eqnarray}
\partial \bar{\partial}\tau &=& 
\frac {2\partial \tau \bar{\partial}\bar{\tau}}
{\bar{\tau}-\tau}\, ,
\label{eomtau8}\\
\partial \bar{\partial} \phi &=& 
\frac{\partial \tau \bar{\partial} \bar{\tau}}{(\tau-\bar{\tau})^2}
(=\partial \bar{\partial}\log \tau_2)\, .
\end{eqnarray}
The first equation eq.(\ref{eomtau8})
is automatically satisfied since we set
$\tau=\tau(z)$.
The second equation is solved as 
\begin{equation}
\phi=\log \tau_2 + F(z) +\bar{F}(\bar{z})
\end{equation}
where $F(z)$ is an arbitrary function of $z$.
In order that eq.(\ref{metric10}) makes sense as a metric,
the field $\phi=\phi(\tau(z),z)$ 
must be  invariant under the modular transformation of $\tau$
and $e^{\phi}$ cannot vanish on anywhere in ${\bf CP}^1$
since we want a solution such that $\tau$ is determined only
up to an $SL(2,{\bf Z})$ transformation.
If $\tau(z)$ is given by eq.(\ref{jtau}) 
with%
\footnote{
When $ord(\Delta)<24$, we can also obtain solutions 
corresponding to eq.(\ref{phi8}).
In this case however, we have to use $w=1/z$ patch 
to indicate the metric around $z\rightarrow \infty$ 
at which the metric becomes singular.
An example of this situation is given in the case of (\ref{e8e8}),
as (\ref{e8z}) and (\ref{e8w}).
} 
\begin{equation}
\Delta= 4 f^3+ 27 g^2 \equiv C \prod_{i=1}^{24}(z-z_i) ,
\end{equation}
these conditions imply
\begin{equation}
\exp(\phi(z, \bar{z}))=\tau_2\eta^2\bar{\eta}^2 
\prod_{i=1}^{24}(z-z_i)^{-\frac {1}{12}}
\prod_{i=1}^{24}(\bar{z}-\bar{z_i})^{-\frac {1}{12}}
\tilde{F}(z)\bar{\tilde{F}}(\bar{z})
\label{phi8}
\end{equation}
where
$\tilde{F}(z)$ is regular and non-vanishing function of $z$
and we take it to be a constant.
The function $\eta$ is Dedekind $\eta$-function :
\begin{equation}
\eta(\tau)=q^{1/24}\prod_{n=1}^\infty(1-q^n) \quad
\left(q=\exp(2\pi i \tau)\right)
\end{equation}
which is needed in eq.(\ref{phi8}) to compensate modular 
transformation of $\tau_2$.
This metric can be extended smoothly to infinity $z\rightarrow\infty$
in ${\bf CP}^1$ since as $z\rightarrow\infty$,
$\tau\rightarrow const.$ and thus $e^\phi\rightarrow (z \bar{z})^{-2}$.

Note that 
\begin{equation}
\tau(z)\sim \frac 1 {2 \pi i}\log(z-z_i) \qquad (z\rightarrow z_i)
\end{equation}
and if we go around $z=z_i$, we have $\tau\rightarrow \tau+1$.
Thus we may say that at $z=z_i$ there is a D7-brane
since it couples to magnetic dual for $\tilde{\phi}(=\tau_1)$ 
field.

The solution (\ref{phi8}) represents important property of 
F-theory compactification
such that at 24 points on the base 
$\tau_2$  goes to infinity and the metric diverges.
Thus 
the existence of such solutions gives us an evidence 
for F-theory.
Or it can be said that the 
solutions support the notion of F-theory.

Note that stringy cosmic string solutions can teach us where
D7-branes are located.
However, they cannot predict symmetry enhancement.
For example, in the F-theory point of view, the elliptic $K3$ with 
$E_8\!\times\! E_8$ 
symmetry is represented as~\cite{mv}
\begin{equation}
y^2=x^3+\alpha z^4 x +z^5+\beta z^6+z^7
\label{e8e8}
\end{equation}
where $\alpha$ and $\beta$ are constants.
In this case, singular fibers of type $E_8$ appear at 
$z=0$ and $z=\infty$
corresponding to the symmetry $E_8\!\times\! E_8$. 
Besides, stringy cosmic string solutions representing eq.(\ref{e8e8})
in the $z$ coordinate system is 
\begin{equation}
\exp(\phi(z, \bar{z}))=
a
\tau_2\eta^2\bar{\eta}^2 
\left|
z^{-\frac {10}{12}} \prod_{i=1}^{4}(z-z_i)^{-\frac {1}{12}}
\right|^2
.
\label{e8z}
\end{equation}
Around $z=\infty$, taking the coordinates $w=1/z$, we have 
\begin{equation}
\exp(\phi)=
\tilde{a}\tau_2\eta^2\bar{\eta}^2 
\left|
w^{-\frac {10}{12}} \prod_{i=1}^{4}(w-\frac 1 {z_i})^{-\frac {1}{12}}
\right|^2\, .
\label{e8w}
\end{equation}
Here $z_i=z_i(\alpha, \beta)$ is defined as
\begin{eqnarray}
\Delta &=& 
z^{10}\left(
4\alpha^3 z^2 +27(1+\beta z + z^2)^2
\right)\\
&\equiv & 27\, z^{10}\prod_{i=1}^{4}(z-z_i) .
\end{eqnarray}
The two representations (\ref{e8z}) and (\ref{e8w}) are transformed to 
each other by usual coordinate transformation 
$z=1/w$ if constants
$a$ and $\tilde{a}$ are suitably chosen.
The above solutions tell us that 
there are respectively ten D7 branes at $z=0$ and $z=\infty$
and they give no explanation of enhanced symmetries.

\section{Stringy cosmic string solutions and F-theory compactifications
on elliptic Calabi-Yau manifolds}
The purpose of this section is to solve the type IIB action (\ref{action})
and find solutions representing F-theory compactified on 
Calabi-Yau three- or four-folds.
Now, consider the compactification of F-theory on elliptic Calabi-Yau
three-folds over two dimensional base $B$.
In the point of view of type IIB theory, this situation 
is represented by solving the equations of motion of the action
(\ref{action}) 
under the assumption 
\begin{equation}
\tau=\tau(z, z')
\end{equation}
where $z=x^8+i x^9$ and $z'=x^6+i x^7$.
The metric is taken to be 
\begin{equation}
ds^2=ds_B^2 + dx_5^2+\cdots dx_1^2 -dt^2 .
\end{equation}
We assume that the metric $ds^2_B$ on $B$ is Hermitian :
\begin{equation}
ds^2_B =g_{z {\bar z}}dz d{\bar z} +g_{z' {\bar z'}}dz' d{\bar z'} +
g_{z {\bar z'}}dz d{\bar z'} +g_{z' {\bar z}}dz' d{\bar z} 
\end{equation}
where 
$$\overline{g_{\alpha\bar{\beta}}}= g_{\beta\bar{\alpha}} . $$
The equation of motion of the action (\ref{action}) 
with respect to $\tau$ 
under the above assumption of 
the metric and the field $\tau$ is
\begin{equation} 
\partial\tau \bar{\partial}g_{z'\bar{z'}}+
\partial'\tau \bar{\partial'}g_{z\bar{z}}-
\partial\tau \bar{\partial'}g_{z'\bar{z}}-
\partial'\tau \bar{\partial}g_{z\bar{z'}}
=0
\label{taueq}
\end{equation} 
where we use the notation
\begin{equation}
\partial \equiv \frac{\partial}{\partial z}, \;
\bar{\partial} \equiv \frac{\partial}{\partial \bar{z}}, \;
\partial' \equiv \frac{\partial}{\partial z'}, \;
\bar{\partial'} \equiv \frac{\partial}{\partial \bar{z'}} \; .
\end{equation}
In contrast to the eight dimensional case, this equation 
gives complicated restriction on 
$\tau=\tau(z,z')$ and on $ds_B^2$.

In order to simplify the problem, we take the ansatz 
$g_{z'\bar{z}}=0$
which corresponds to taking a diagonal metric.
We write 
\begin{eqnarray}
g_{z\bar{z}}&=& e^{\phi(z,\bar{z},z',\bar{z'})} \, ,\\
g_{z'\bar{z'}}&=& e^{\psi(z,\bar{z},z',\bar{z'})}.
\end{eqnarray}
To meet the eq.(\ref{taueq}) independent of $\tau$,
we further assume 
\begin{equation}
\phi = \phi(z,\bar{z}),\quad \psi = \psi(z',\bar{z'}) .
\end{equation}
Next, we have to solve Einstein equations:
\begin{equation}
R_{\mu\nu}-\frac 1 2 R g_{\mu\nu}=T_{\mu\nu}
\end{equation}
with
\begin{equation}
T_{\mu\nu}= \frac 1 2 \frac {\partial_{(\mu}\tau\partial_{\nu)}\bar{\tau}}
{\tau_2{}^2}
-\frac 1 4 
\frac {\partial_{\rho}\tau g^{\rho\sigma}\partial_{\sigma}\bar{\tau}}
{\tau_2{}^2}g_{\mu\nu}.
\end{equation}
Under the assumptions we have made, the Einstein equations reduce to
the following three equations :
\begin{eqnarray}
0 &=& \frac{\partial\tau \bar{\partial'}\bar{\tau}
\pm \partial'\tau \bar{\partial}\bar{\tau}}
{\tau_2^2}\, ,\\
\partial\bar{\partial}\phi &= &
\frac{\partial\tau\bar{\partial}\bar{\tau}}{\tau_2^2}
\label{eq2}\, ,
\\
\partial'\bar{\partial'}\phi &= &
\frac{\partial'\tau\bar{\partial'}\bar{\tau}}{\tau_2^2} 
\label{eq3}.
\end{eqnarray}
The first equation leads
\begin{equation}
\partial\tau= 0 \quad {\rm or} \quad \partial'\tau=0\, ,
\end{equation}
and we take $\partial'\tau =0$, i.e., $\tau =\tau(z)$.
Then, eqs.(\ref{eq2}) and (\ref{eq3}) can be solved as
\begin{eqnarray}
\phi &=& \log\tau_2 + F(z) + \bar{F}(\bar{z})\, ,\\
\psi &=& F'(z) + \bar{F'}(\bar{z})\, .
\end{eqnarray}
Using the similar arguments as in the case of eight dimensions,
\begin{eqnarray}
\exp(\phi(z, \bar{z}))&=&
a\tau_2\eta^2\bar{\eta}^2 
\left|
\prod (z-z_i)^{-\frac {1}{12}}
\right|^2
\label{tau61}\\
\exp(\psi(z', \bar{z'}))&=& |F'(z')|^2
\label{tau62}
\end{eqnarray}
where $z_i$ represents zero point of $\Delta(z, z')$,
i.e., $\Delta(z, z')= C(z')\prod (z-z_i)$ and 
$F'(z')$ is a non-vanishing function.

Now let us consider 
what happens when we restrict
$\tau =\tau(z)$ 
in the F-theory point of view.
As is described in section 2, 
if the base $B={\bf F}_n$,
polynomials 
$$f(z,z')= \sum_{k=-4}^{4}f_{8-nk}(z')z^{4-k}$$ 
and 
$$g(z,z')=\sum_{k=-6}^{6}g_{12-nk}(z')z^{6-k}$$
determine the structure of elliptic fiber $\tau$ as
in eq.(\ref{jtau}).
There are the following three cases satisfying
$\tau=\tau(z)$:
\begin{eqnarray}
&{\rm (a)}& \qquad 
\left\{
\begin{array}{l}
f(z,z')= f_0(z) h(z')^2\\
g(z,z')=g_0(z) h(z')^3
\end{array}
\right.
\\
&{\rm (b)}& \qquad f(z,z')=0 \\
&{\rm (c)}& \qquad g(z,z')=0 .
\end{eqnarray}
The last two cases (b) and (c) correspond to 
the constant coupling solutions 
$\tau= const.,$ and we discuss them later on.
Here we consider the case (a).
In this case the order of $f_0(z)$, $g_0(z)$ and $h(z')$ depend on 
the base $B$ we choose.
The discriminant is 
\begin{equation}
\Delta= h(z')^6 \left( 4 f_0(z)^3+27 g_0(z)^2
\right) .
\end{equation}

The twelve dimensional metric is naturally taken to be
the following modular invariant form~\cite{tse}:
\begin{equation}
ds_{12}= h_{pq}dy^pdy^q +ds_B + ds_{6{\cal M}}
\end{equation}
where  
$ds_{6{\cal M}}$ denotes the six dimensional Minkowski metric,
$y^p$ the coordinates of the fiber torus 
and
\begin{equation}
h_{pq}=\frac{1}{\tau_2}
\left(
\begin{array}{cc}
|\tau| ^2& \tau_1 \\
\tau_1 & 1
\end{array}
\right) .
\end{equation}
Note however, that 
in order to obtain the theory with $N\!=\!1$ supersymmetry in 
six dimensions,
we should modify this metric 
such as internal metric on the elliptic Calabi-Yau manifold 
to be a Ricci-flat K\"{a}hler metric \cite{gsvy}.

In the following, we consider the case 
$B={\bf CP}^1\!\times\!{\bf CP}^1(={\bf F}_0)$ and ${\bf F}_n$, respectively.

\subsection{$B={\bf CP}^1\times {\bf CP}^1$}
Now we consider the case of $B={\bf CP}^1\!\times\! {\bf CP}^1$.
The coordinates $z$ and $z'$ are 
naturally taken to be 
those of two ${\bf CP}^1$'s.
{}From eq.(\ref{fn}),
the polynomial $f(z,z')$ (or $g(z,z')$) is generically
of order 8 (or 12) in each of $z$ and $z'$.
If 
it is the case of (a),
$f_0$ and $g_0$ are of order 8 and 12 in $z$ respectively,
and $h$ is of order 4 in $z'$.
Thus the discriminant and $\tau=\tau(z)$ are given explicitly as 
\begin{eqnarray}
\Delta &=& \{4 f_0(z)^3 + 27 g_0(z)^2\} h(z')^6\\
&\equiv & C \prod_{i=1}^{24}(z-z_i) 
\prod_{i=1}^{4}(z'-z'_i)^6 
\end{eqnarray}
and
\begin{equation}
J(\tau(z))= \frac{4 \left(24 f_0(z)\right)^3}{4 f_0(z)^3 + 27 g_0(z)^2} \, .
\label{p1p1j}
\end{equation}
We see that the elliptic fiber degenerates on  codimension 1 locus
 $\Delta=0$ on $B$, i.e.,
on $z=z_i$ $(i=1,\cdots 24)$ and on $z'=z'_i$ $(i=1,\cdots 4)$.
On the other hand, 
$\tau$ diverges on $z=z_i$ $(i=1,\cdots , 24)$ 
which correspond to parts of D7-brane worldvolume.

Specializing (\ref{tau61}) and (\ref{tau62}) to this case,
we naturally obtain 
\begin{eqnarray}
e^{\phi} &=&
a\tau_2\eta^2\bar{\eta}^2 
\prod_{i=1}^{24}(z-z_i)^{-\frac {1}{12}}
\prod_{i=1}^{24}(\bar{z}-\bar{z_i})^{-\frac {1}{12}}\, ,
\label{ppphi}\\
e^{\psi} &=&
a' \prod_{i=1}^{4}(z'-z'_i)^{-\frac {1}{2}}
\prod_{i=1}^{4}(\bar{z'}-\bar{z'_i})^{-\frac {1}{2}} .
\label{pppsi}
\end{eqnarray}
By the similar argument as in the case of eight dimensions,
we see that both $e^{\phi}dz d\bar{z}$ and $e^{\psi}dz' d\bar{z'}$
represent the metric on ${\bf CP}^1$ globally
since 
$e^\phi\rightarrow (z\bar{z})^{-2}$ as $z\rightarrow \infty$ and 
$e^\psi\rightarrow (z'\bar{z'})^{-2}$ as $z'\rightarrow \infty$.
Therefore we conclude that we obtain the cosmic string solutions 
representing F-theory compactified on elliptic Calabi-Yau manifold
over the base  
$B={\bf CP}^1\!\times\! {\bf CP}^1\ni \{z,z'\}$ with $\tau = \tau (z)$.
It can also be said that we have given explicit examples of 
compact D-manifolds~\cite{bvs}.

Note that the above argument can easily be extended to the case
of an elliptic Calabi-Yau four-fold over the base 
$B={\bf CP}^1\!\times\! {\bf CP}^1\!\times\! {\bf CP}^1\ni \{z,z', z''\}$ 
with $\tau = \tau (z)$.

\subsection{$B={\bf F}_n$}

Various properties of F-theory compactified on 
elliptic Calabi-Yau manifolds with $B={\bf F}_n$ 
such that spectrum or enhanced gauge symmetry 
have been closely investigated~\cite{mv,ber}.
We thus want to construct stringy cosmic string solutions
reflecting these properties.
We describe the results by taking an example $B={\bf F}_4$.
Other cases can be treated similarly.

We know from the structure of singular fiber on $B={\bf F}_4$ 
that the theory 
has generically symmetry enhancement of ${\rm SO}(8)$.
The elliptic modulus $\tau(z,z')$ is given by eq.(\ref{fn})
with $n=4$:
\begin{equation}
J(\tau(z,z'))= \frac{ 4\left( 24 f(z,z')\right)^3}{4 f(z,z')^3+ 27 g(z,z')^2}
\end{equation}
\begin{eqnarray}
f(z,z' ) &=& 
z^2 \sum_{k=0}^{6}  f_{4k}(z')z^{k}
\\
g(z,z' ) &=& 
z^3 \sum_{k=0}^{9} g_{4k}(z')z^{k} 
\end{eqnarray}
where $f_{4k}$ ( or $g_{4k}$) is as usual a polynomial of 
order $\le 4k$ in $z'$.
Thus generically we may write  
\begin{equation}
\Delta= C z^6 \prod_{i=1}^{18} \{ z+ h_i(z') \}\, .
\end{equation}
Here  $h_{i}$ is of order $\le 4$.
We restrict $\tau$ to depend only on one coordinate $y=y(z,z')$ in order that 
we can solve equations of motion of the action (\ref{action}) easily.
Choosing an order $\le 4$ function $\xi(z')$, we take $y=z+\xi(z')$ 
and $\tau=\tau(z+\xi(z'))$.
This restriction corresponds to taking 
\begin{equation}
f=\tilde{f}_6(y)z^2\, , \quad g=\tilde{g}_9 (y)z^3 .
\end{equation}
By using a coordinate system $(y,y')= (z+\xi(z'),z')$,
we assume the metric on $B$ as 
\begin{equation}
ds^2_{B}= e^{\phi(y,\bar{y},y',\bar{y'})} 
dyd\bar{y}+e^{\psi(y,\bar{y},y',\bar{y'})} dy' d\bar{y'}.
\end{equation}
The fields $\phi$ and $\psi$ can be determined as the same way
as in the case $B={\bf CP}^1\!\times\! {\bf CP}^1$:
\begin{eqnarray}
\exp(\phi(y,\bar{y}))&=&
a \tau_2\eta^2\bar{\eta}^2 
\left|
\prod_{i=1}^{18} (y-y_i)^{-\frac {1}{12}}
\right|^2
\\
\exp(\psi(y', \bar{y'}))&=& |G(y')|^2
\end{eqnarray}
where 
\begin{eqnarray}
\Delta = C z^6 \prod_{i=1}^{18} (y-y_i) .
\end{eqnarray}
Coordinate transformation back to $(z,z')$ leads
\begin{equation}
ds^2_{B}= e^{\phi}dz d\bar{z} + 
e^\phi\left(\frac{\partial \xi}{\partial z'} dz'd\bar{z}+ c.c.\right)
+
\left(e^\phi \left|\frac{\partial \xi}{\partial z'} \right|^2
+ e^{\psi} \right) dz'd\bar{z'} .
\end{equation}
Furthermore, we have to 
take into account the effect of zero locus 
$z=0$ of $\Delta$
corresponding to the symmetry $SO(8)$ ($D_4$ singularity) on $z=0$.
The result is 
\begin{equation}
ds^2_{B}= e^{\phi}|z|^{-1} dz d\bar{z} + 
e^\phi\left(\frac{\partial \xi}{\partial z'} \bar{z}^{-\frac 1 2}
dz'd\bar{z}+ c.c.\right)
+
\left(e^\phi \left|\frac{\partial \xi}{\partial z'} \right|^2
+ e^{\psi} \right) dz'd\bar{z'} .
\label{f4dsb}
\end{equation}
We can check that this metric certainly satisfy 
Einstein equations.
This solution represents that $\tau_2$ diverges on $y=y_i$
($y_i\!=\!1,\cdots 18$)
and the metric diverges on $\Delta=0$.
These are the properties expected from F-theory 
point of view.
If we fix a point $z'$ on the base of ${\bf F}_4$ and 
take the limit $z\rightarrow \infty$,
the metric becomes 
$$
e^\phi|z|^{-1} dzd\bar{z} \rightarrow (z \bar{z})^{-2}dzd\bar{z} .
$$
Thus this solution seems to represent that the fiber of ${\bf F}_4$
is ${\bf CP}^1$ globally.
However,
note that contrary to the $B={\bf CP}^1\! \times\!{\bf CP}^1$ case,
this solution is only applicable in the region $|z'|<\infty$.
We do not know the way of extending this solution 
to cover all regions of ${\bf F}_4$.

For other cases $B={\bf F}_n$  we can obtain the similar solution 
if $\tau$ is restricted as $\tau=\tau(z+\xi_n(z'))$
where $\xi_n(z')$ is a polynomial of order $\le n$.

\subsection{Constant coupling solutions}
It is known that a
constant coupling limit $\tau(z)\rightarrow const.$ 
(up to an $SL(2,{\bf Z})$ transformation) of F-theory
compactified on $K3$ is
corresponding to orbifold limit of 
$K3\rightarrow T^4/{\bf Z}_n$ with $n=2,3,4$ or $6$~\cite{sen,dm}.
Similar considerations in the case of 
elliptic Calabi-Yau three-fold over $B={\bf CP}^1\! \times\!{\bf CP}^1$
have been done in ref.\cite{an,dp,sen2}.
More generally, the relation between
F-theory compactified on Calabi-Yau $n$-folds and 
type IIB orientifold theories 
have been investigated \cite{sen3}.
Here we consider F-theory compactifications on 
Calabi-Yau three-folds with $\tau$ remains constant
in the point of view of stringy cosmic string solutions.

First, we consider the case $B={\bf CP}^1\! \times\!{\bf CP}^1$.
In this case, since we already have solutions with 
$\tau=\tau(z)$  in section 4.1, we can obtain constant 
coupling solutions by taking a certain limit of them.
There are the following three cases of obtaining 
constant coupling solutions:
\begin{eqnarray}
&{\rm (a)} &f_0(z)\rightarrow \phi(z)^2\, ,
\, g_0(z)\rightarrow \alpha \phi(z)^3\\
&{\rm (b)}& g_0(z)\rightarrow 0 \\
&{\rm (c)}&f_0(z)\rightarrow 0 .
\end{eqnarray}
They are respectively corresponding to the cases
(a), (b) and (c) described in the first part of this section.
Note that in the first case the value of $\tau$ is determined by the 
constant $\alpha$ whereas  
$\tau=i$ in the case (b)
and $\tau=e^{i\pi/3}$ in (c).
In the first case, we see from (\ref{ppphi}) and (\ref{pppsi})
that 
the metric on the base $B$ becomes
\begin{equation}
ds^2_B=
a
\prod_{i=1}^{4}(z-z_i)^{-\frac {1}{2}}
\prod_{i=1}^{4}(\bar{z}-\bar{z_i})^{-\frac {1}{2}}dz d\bar{z}
+ a'
\prod_{i=1}^{4}(z'-z'_i)^{-\frac {1}{2}}
\prod_{i=1}^{4}(\bar{z'}-\bar{z'_i})^{-\frac {1}{2}} dz' d\bar{z'}.
\end{equation}
We see that the topology of $B$ is considered as 
$T^2/{\bf Z}_2\times T^2/{\bf Z}_2$. It is claimed 
in ref.\cite{ja,an} that 
this limit is related to 
F-theory compactified on $T^6/({\bf Z}_2\!\times\!{\bf Z}_2)$.
Similarly, in the remaining cases, we can specify the metric on $B$
and they can be interpreted as orbifold limit of 
F-theory compactifications.

Besides,  
independent of above three cases, we can construct several
constant coupling solutions.
For example, the metric 
\begin{equation}
ds^2_B=
a \prod_{i=1}^{3}(z-z_i)^{-\frac {2}{3}}
\prod_{i=1}^{3}(\bar{z}-\bar{z_i})^{-\frac {2}{3}}dz d\bar{z}
+
a' \prod_{i=1}^{3}(z'-z'_i)^{-\frac {2}{3}}
\prod_{i=1}^{3}(\bar{z'}-\bar{z'_i})^{-\frac {2}{3}} dz' d\bar{z'}
\end{equation}
corresponds to a solution with $\tau=e^{i\pi/3}$
and it is considered to be related to the 
F-theory compactified on $T^6/({\bf Z}_3\!\times\!{\bf Z}_3)$.

In the case of $B={\bf F}_4$, taking 
$f(z,z')\rightarrow \phi(z,z')^2$ and $g(z,z')\rightarrow \alpha \phi(z,z')^3$
in eq.(\ref{f4dsb}), the metric becomes
\begin{eqnarray}
ds^2_{(B)} &\rightarrow &
\left| \prod_{i=1}^{4}\left(z+\xi(z')-y_i\right)^{-\frac 1 2} 
 \right|^2
|z|^{-1} dz d\bar{z} \nonumber\\
& &+ 
\left| \prod_{i=1}^{4}\left(z+\xi(z')-y_i\right)^{-\frac 1 2}
 \right|^2
\left(\frac{\partial \xi}{\partial z'} \bar{z}^{-\frac 1 2}
dz'd\bar{z}+ c.c.\right)\nonumber\\
& &
+ \left(\left| \prod_{i=1}^{4}\left(z+\xi(z')-y_i\right)^{-\frac 1 2}
\frac{\partial \xi}{\partial z'} \right|^2
+ |G(z')|^2\right) dz'd\bar{z'} .
\end{eqnarray}
In this case we have no information of determining the form of 
$G(z')$ and 
we cannot conclude that 
this limit corresponds to the orbifold limit of F-theory compactification
at least from the point of view of stringy cosmic string solutions.

\subsection{$h^{1,1}=h^{2,1}=19$ model}
We look for stringy cosmic string solutions corresponding to
the $h^{1,1}\!=\!h^{2,1}\!=\!19$ 
Calabi-Yau manifold described in eq.(\ref{cy19}).
This manifold is represented as an elliptic fibration over 
a base $B$.
However, as we described in section 2, it is also interpreted as
two equivalent tori $T$ and $\tilde{T}$
are fibered over the base ${\bf CP}^1$.
One of the two tori $T$ is identified with the elliptic fiber
over $B$ and the other $\tilde{T}$ is 
considered to be embedded in the base $B$.
These fiber tori are explicitly represented as 
\begin{eqnarray}
y^2 &=& x^3+f(z)x +g(z) ,
\label{t1}\\
\tilde{y}^2 &=& \tilde{x}^3+\tilde{f}(z)\tilde{x} +\tilde{g}(z)
\label{t2}
\end{eqnarray}
where $(x,y)$ and $(\tilde{x},\tilde{y})$
denote the tori $T$ and $\tilde{T}$ respectively.
The order of polynomials $f(z)$ and $\tilde{f}(z)$ is $\le 4$
and that of $g(z)$ and $\tilde{g}(z)$ is $\le 6$.
The discriminant of (\ref{t1}) and (\ref{t2})
are respectively
\begin{eqnarray}
\Delta &=& 4 f^3+ 27 g^2 , \\
\tilde{\Delta} &=& 4 \tilde{f}^3+ 27 \tilde{g}^2 .
\end{eqnarray}
The torus $T$ (or $\tilde{T}$) degenerates on 
$\Delta=0$ ( or $\tilde{\Delta}=0$).

Now we look for stringy cosmic string solutions corresponding to 
the above situation.
We start with the type IIB effective action in ten  dimensions as before :
\begin{equation}
S_{10}= -\frac 1 2\int d^{10}x \sqrt{g_{(10)}}
\left( R_{(10)} -\frac 1 2 \frac {\partial_\mu\tau g^{\mu\nu}\partial_\nu
{\bar \tau}}
{\tau_2^2} \right).
\end{equation}
We further compactify this on torus $\tilde{T}$.
If we assume that the volume of two tori $T$ and $\tilde{T}$ 
is both fixed to 1, the  resulting eight dimensional action 
after dimensional reduction becomes 
\begin{equation}
S_{8}= 
-\frac 1 2
 \int d^{8}x
\sqrt{g_{(8)}}
\left( R_{(8)} -\frac 1 2 \frac {\partial_\mu\tau g^{\mu\nu}\partial_\nu
{\bar \tau}}{\tau_{2}^2} 
- \frac 1 2 \frac {\partial_\mu\tilde{\tau} g^{\mu\nu}\partial_\nu
{\bar{\tilde{\tau}}}}{\tilde{\tau}_2^2} 
\right).
\label{action19}
\end{equation}
where we use 
\begin{equation}
h_{ij}=
\frac 1{\tau_2}
\left(
\begin{array}{cc}
|\tau|^2& \tau_1 \\
\tau_1 & 1
\end{array}
\right) 
\end{equation}
as a metric on torus $\tilde{T}$.

In order to obtain solutions of the action, we take the ansatz
on the eight dimensional metric as
\begin{equation}
ds^2_{(8)}=e^{\phi(z,\bar{z})}dz d\bar{z} + 
ds^2_{6{\cal M}}
\end{equation}
where the coordinate  $z=x_6+ i x_7$ 
represents the base ${\bf CP}^1$.
The total twelve dimensional metric is considered as 
\begin{equation}
ds^2{}_{12}=h_{pq}dy^pdy^q +  \tilde{h}_{pq}d\tilde{y}^pd\tilde{y}^q +
  ds^2{}_{(8)} 
\end{equation}
where 
\begin{equation}
h_{pq}=\frac 1{\tau_2}
\left(
\begin{array}{cc}
|\tau|^2& \tau_1 \\
\tau_1 & 1
\end{array}
\right) \quad ,\quad
\tilde{h}_{pq}=\frac 1{\tilde{\tau}_2}
\left(
\begin{array}{cc}
|\tilde{\tau}|^2& \tilde{\tau}_1 \\
\tilde{\tau}_1 & 1
\end{array}
\right) .
\end{equation}
The equations of motion  of the action (\ref{action19}) with respect to 
$\tau $ and $\tilde{\tau}$ have similar form as eq.(\ref{eomtau8})
and they are satisfied by taking 
$\tau=\tau(z)$ and $\tilde{\tau}=\tilde{\tau}(z)$.
The only independent equation in
the  Einstein equations is
\begin{eqnarray}
\partial \bar{\partial}\phi &=&
\frac{\partial \tau \bar{\partial} \bar{\tau}}{(\tau-\bar{\tau})^2}
+
\frac{\partial \tilde{\tau} \bar{\partial} \bar{\tilde{\tau}}}
{(\tilde{\tau}-\bar{\tilde{\tau}})^2}\\
&=&
\partial \bar{\partial}\left(
\log\tau_2 +\log\tilde{\tau}_2
\right)
\end{eqnarray}
which is solved as 
\begin{equation}
\phi= \log\tau_2 +\log\tilde{\tau}_2 + F(z) + \bar{F}(\bar{z}) .
\end{equation}
As in the previous cases,
the field 
$\phi$ must be invariant under modular transformations 
of both $\tau$ and $\tilde{\tau}$, and 
$e^\phi$ must be nonzero everywhere on ${\bf CP}^1$.
Thus,
\begin{equation}
\exp(\phi(z, \bar{z}))=
a \tau_2\eta^2\bar{\eta}^2 
\tilde{\tau}_2\tilde{\eta}^2\bar{\tilde{\eta}}^2 
\left|
\prod_{i=1}^{12} (z-z_i)^{-\frac {1}{12}}
\right|^2
\left|
\prod_{j=1}^{12} (z-y_j)^{-\frac {1}{12}}
\right|^2 .
\end{equation}
Here
$\eta\equiv\eta(\tau)$ and $\tilde{\eta}\equiv\eta(\tilde{\tau})$
and $z_i$ and $y_i$ are given by 
\begin{eqnarray}
\Delta(z) &\equiv & C\prod_{i=1}^{12}(z-z_i) \, ,\\
\tilde{\Delta}(z)&\equiv &  \tilde{C}\prod_{j=1}^{12}(z-y_j) .
\end{eqnarray}
Since generically the order of $\Delta$ or $\tilde{\Delta}$
is 12, at infinity we have
\begin{equation}
e^{\phi(z, \bar{z})}dz d\bar{z}\rightarrow  \frac{1}{(z\bar{z})^2}dz d\bar{z}
\qquad (|z|\rightarrow \infty) \, .
\end{equation}
Thus this metric indicates that the topology of the eight 
dimensional spacetime is certainly ${\bf CP}^1\times M^6$.
Each of $\Delta=0$ locus $z=z_i$ corresponds to D7-brane
worldvolume.

\section{Conclusions} 
In this paper, we have solved equations of motion of 
low energy effective action of 
ten dimensional type IIB theory and have obtained
stringy cosmic string solutions 
corresponding to compactifications of F-theory on 
elliptic Calabi-Yau 3,4-folds. 
In concrete, we have constructed solutions 
in the case of $B=({\bf CP}^1)^{\otimes n}$, ${\bf F}_n$ and a blown-up 
${\bf CP}^2$,
and have investigated their constant coupling limits.
For each of these manifolds, our solutions do not cover 
whole moduli space of the F-theory compactifications,
which is caused by technical reasons.
However, solutions corresponding to more general cases should exist.

Existence of such solutions is considered as one of the
evidences of F-theory compactifications.
In particular, the solutions
explicitly show the 
codimension 1 surface in the base $B$ representing 
the locations of D7-branes.

\paragraph{Acknowledgments}
The author would like to thank M. Natsuume for useful discussions.
This work is supported by 
Japan Society for the Promotion of Science. 

\end{document}